\documentclass[prb, twocolumn,amsmath,amssymb]{revtex4}
\usepackage{graphicx, color}
\usepackage{dcolumn}
\usepackage{bm}
\begin{document}
\title{Correlated spin currents generated by resonant crossed Andreev reflections in topological superconductors}
\author{James J. He$^{1}$, Jiansheng Wu$^{1}$, Ting-Pong Choy$^{1,2}$, Xiong-Jun Liu$^{1,2}$, Y. Tanaka$^{3}$}
\author{ K. T. Law$^{1}$}\thanks{Correspondence address : phlaw@ust.hk}
\affiliation{$^{1}$ Department of Physics, Hong Kong University of Science and Technology, Clear Water Bay, Hong Kong, China}
\affiliation{$^{2}$ Institute for Advanced Studies, Hong Kong University of Science and Technology, Clear Water Bay, Hong Kong, China}
\affiliation{$^{3}$ Department of Applied Physics, Nagoya University, Nagoya, Japan}
\date{\today}
\begin{abstract}
{\bf Topological superconductors, which support Majorana fermion excitations, have been the subject of intense studies due to their novel transport properties and their potential applications in fault-tolerant quantum computations. Here we propose a new type of topological superconductors which can be used as a novel source of correlated spin currents. We show that inducing superconductivity on a AIII class topological insulator wire, which respects a chiral symmetry and supports protected fermionic end states, will result in a topological superconductor. This topological superconductor supports two topological phases with one or two Majorana fermion end states respectively. In the phase with two Majorana fermions, the superconductor can split Cooper pairs efficiently into electrons in two spatially separated leads due to Majorana induced resonant crossed Andreev reflections. The resulting currents in the leads are correlated and spin-polarized. Importantly, the proposed topological superconductors can be realized using quantum anomalous Hall insulators in proximity to superconductors.}
\end{abstract}

\maketitle
\section{\bf Introduction}
The search for topological superconductors which support Majorana fermions (MFs) [\onlinecite{Wilczek}] has attracted much theoretical and experimental studies in recent years [\onlinecite{KH, Moore, QZ, Alicea, Beenakker, Franz, ST}]. These studies are strongly motivated by the fact that MFs in topological superconductors are non-Abelian particles and have potential applications in fault-tolerant quantum computations [\onlinecite{Kitaev, NSSFS}]. Recent studies have pointed out that one of the most promising ways to engineer topological superconductors is by inducing s-wave superconductivity on semiconductor wires with Rashba spin-orbit coupling in the presence of external magnetic fields [\onlinecite{STF, SLTD, LSD, Alicea2, ORV, BDRv, PL11}]. This results in so-called D class topological superconductors which break time-reversal symmetry and support a single MF end state at each end of a superconducting wire [\onlinecite{SRFL, TK}]. These D class topological superconductors also exhibit a number of interesting transport properties such as fractional Josephson effects [\onlinecite{Kitaev2, Kwon}], resonant Andreev reflections [\onlinecite{LLN, WADB}], enhanced crossed Andreev reflections [\onlinecite{Nilsson, Jie}]. So far, the search for D class topological superconductors has been one of the most important areas in the study of topological superconductors.

However, according to the Altland-Zirnbauer symmetry classification scheme [\onlinecite{SRFL}], there exist other topological superconductors which belong to different symmetry classes. Many aspects of the physical properties and potential applications of various types of topological superconductors have yet to be explored. 

\begin{figure}
\begin{center}
\includegraphics[width=3.2in]{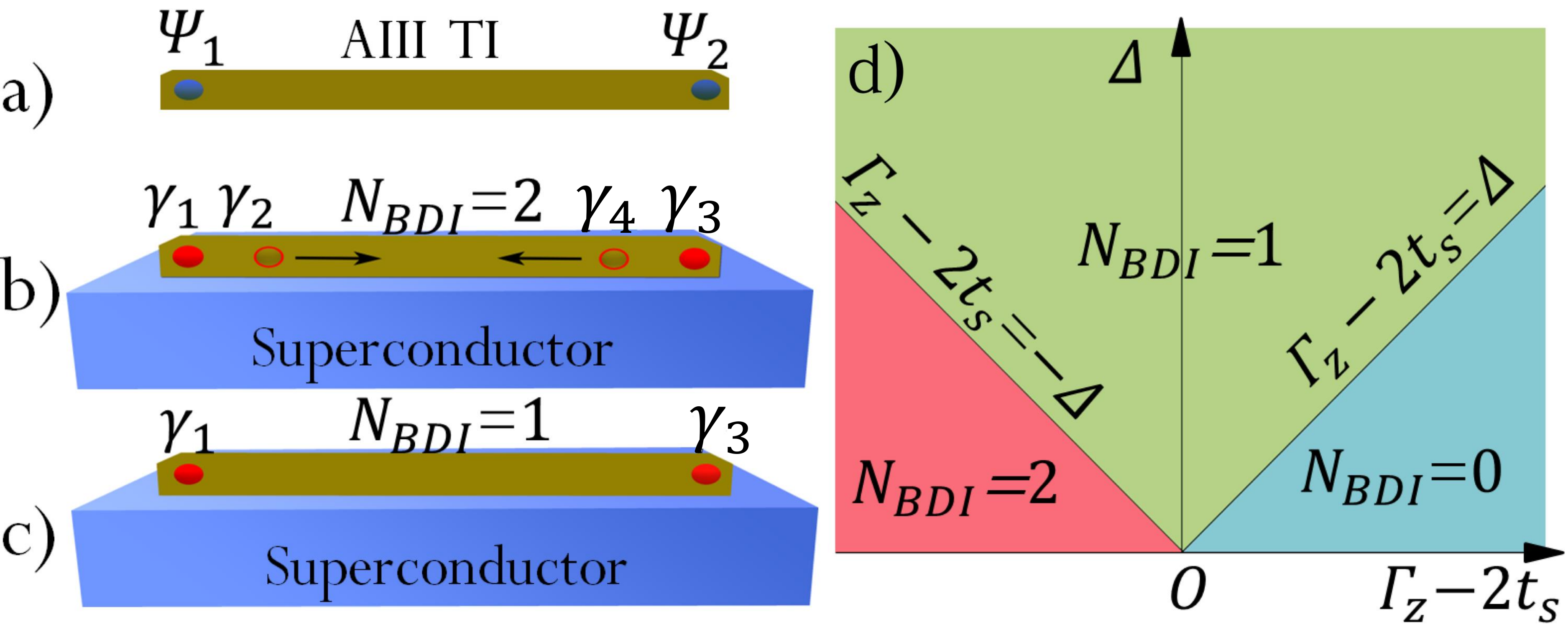}
\caption{ From a AIII class topological insulator to a BDI topological superconductor. (a) AIII topological insulator with fermionic end states $\Psi_{1}$ and $\Psi_2$ located at the ends of the wire. (b) By inducing superconductivity on the AIII class topological insulator, each fermionic end state becomes two MFs and the system becomes a BDI class topological superconductor in the $N_{\rm BDI}=2$ phase. As the pairing strength $\Delta$ increases, one of the Majorana fermions at each end merges into the bulk. (c) After the bulk gap is closed by increasing $\Delta$, only one MF end state is left at each end and we have the $N_{\rm BDI}=1$ phase. (d) Phase diagram of the BDI class topological superconductor characterized by the topological invariant $N_{\rm BDI}$ as functions of the pairing strength $ \Delta$ and $\Gamma_z-2t_{\rm so}$.}
\label{Fig1}
\end{center}
\end{figure}

Here, we demonstrate that  inducing s-wave superconductivity on a AIII class topological insulator [\onlinecite{SRFL, TK}], which respects a chiral symmetry and supports fermionic end states as illustrated in Fig.1a, will result in a new type of superconductor. The resulting superconductor is in the BDI class, which respects a time-reversal like symmetry and the particle hole symmetry and it is classified by an integer topological invariant $N_{\rm BDI}$ [\onlinecite{TK, Tewari, Chris, Sato}]. This BDI class superconductor supports two distinct topological phases distinguished by one ($N_{\rm BDI}=1$) and two ($N_{\rm BDI}=2$) MF end states at each end of the wire respectively. While the superconductor in the $N_{\rm BDI}=1$ phase has almost identical transport properties as a D class topological superconductor, the $N_{\rm BDI}=2$ phase exhibits several transport anomalies. Particularly, in the $N_{\rm BDI}=2$ phase, local Andreev reflections (ARs) are completely suppressed at the normal lead/topological superconductor interface at zero bias due to the destructive interference between the AR amplitudes induced by the two MFs. When two normal leads are attached to the two ends of the superconductor, resonant crossed AR processes can happen, causing an electron from one normal lead to be reflected as a hole in the other lead with probability of unity. In reverse processes, when a current is driven from the superconductor to the leads, Cooper pairs can split into two spatially separated leads and form correlated electron pairs with perfect efficiency. We call this phenomenon resonant Cooper pair splitting. Remarkably, the outgoing currents of the two leads are correlated and spin-polarized. Importantly, we show that these unique transport properties of BDI class topological superconductors can be experimentally realized using quantum anomalous Hall insulators in proximity to an $s$-wave superconductor. 

\section{\bf Results}

\subsection{ From class AIII to class BDI }

In this section, we first point out how to obtain a BDI class topological superconductor from an AIII class topological insulator. The properties of the MF end states are also studied. Secondly, we study the local AR properties of the BDI class topological superconductor by attaching a normal lead to one end of the topological superconductor. Thirdly,  we examine the effects of resonant crossed ARs and resonant Cooper pair splitting induced by the double MF end states in the $N_{\rm BDI}=2$ phase. The generation of correlated spin currents using these novel phenomena is also discussed. Lastly, we discuss the relation between the BDI class topological superconductor and quantum anomalous Hall insulators.

An AIII class topological insulator is  a one dimensional system which respects a chiral symmetry and supports fermionic end states [\onlinecite{TK, XJ}]. A simple AIII class Hamiltonian, which can be topologically non-trivial in the basis of $(c_{k \uparrow},c_{k \downarrow})$, can be written as [\onlinecite{XJ}]
\begin{equation}
H_{\rm AIII}(k) = (\Gamma_z-2 t_{\rm s}\cos k ) \sigma_z+ 2 t_{\rm so}\sin k \sigma_y .
\end{equation}
Here, $c_{k \uparrow}$ ($c_{k \downarrow}$) denotes a spin up (down) fermionic operator, $t_{\rm s}$ is the hopping amplitude, $\Gamma_{z}$ is the Zeeman term and $t_{\rm so} $ is the hopping amplitude with spin flip. For simplicity and without loss of generality, we assume $t_{\rm s}$, $t_{\rm so}$ and $\Gamma_{z}$ to be positive real numbers. Since the Hamiltonian contains only the $\sigma_{y}$ and $\sigma_{z}$ terms, $H (k)$ respects the chiral symmetry $\sigma_{x} H(k) \sigma_{x}=-H(k)$ and $H(k)$ belongs to AIII class according to symmetry classifications [\onlinecite{SRFL}]. In the regime where $|\Gamma_{z}|< 2t_{\rm s}$, $H(k)$ is topologically non-trivial. For a topologically non-trivial AIII class wire with open boundaries, the wire supports a single fermionic end state at each end of the wire  [\onlinecite{XJ}] as depicted in Fig.1a.

Interestingly, the AIII class topological insulator becomes a BDI class topological superconductor when superconducting $s$-wave pairing terms $\Delta_{0} {c}_{k \uparrow} {c}_{-k \downarrow} + \text{H.c.} $ are added. 
In the Nambu basis  $(c_{k \uparrow},c_{k\downarrow},c_{-k\uparrow}^{\dag},c_{-k\downarrow}^{\dag})$ the Hamiltonian is:
\begin{equation}
H_{\rm BDI}(k) = \left[ \left(\Gamma_z-2 t_{\rm s}\cos k\right) \sigma_z+ 2 t_{\rm so}\sin k\sigma_y \right]\tau_z +\Delta \sigma_y \tau_y, \ \
\end{equation}
where $\sigma_{i}$ and $\tau_{i}$ are Pauli matrices acting on spin and particle-hole space, respectively.

In the presence of the pairing terms, the symmetry class of the Hamiltonian is changed from AIII to BDI. In particular, we note that the Hamiltonian satisfies a time-reversal like symmetry $ {\mathcal T}H_{\rm BDI}({k}) {\mathcal T}^{-1}= H_{\rm BDI}({-k})$ and a particle-hole symmetry $ {\mathcal P} H_{\rm BDI}(k) {\mathcal P}= -H_{\rm BDI}(-k)$, where $\mathcal{T}=\sigma_x\tau_x  \mathcal{K}$, $\mathcal{P}=\sigma_{0}\tau_x \mathcal{K}$ and $\mathcal{K}$ is the complex conjugate operator. Since $\mathcal{T}^2=1$, there is no Kramer's degeneracy associated with $\mathcal{T}$. As a result of $\mathcal{T}$ and $\mathcal{P}$ symmetries, we have $\mathcal{C}H_{\rm BDI}(k) \mathcal{C}^{-1}=-H_{\rm BDI}(k)$, where $\mathcal{C}=\mathcal{TP}=\sigma_{x}\tau_{0}$. Therefore, $H_{\rm BDI}(k)$ is in the BDI class [\onlinecite{SRFL, TK}]. It has been shown that a BDI class topological superconductor is classified by an integer topological invariant $N_{\rm BDI}$ [\onlinecite{SRFL, TK, Tewari, Chris, Sato}], which denotes the number of topologically protected MF end states at each end of the superconducting wire.

The topological invariant $N_{\rm BDI}$ can be easily evaluated [\onlinecite{Chris}] and the phase diagram of $H_{\rm BDI}$, as functions of $\Gamma_z - 2t_s$ and $\Delta$, is depicted in Fig.1d. It is evident that there are two topological phases with $N_{\rm BDI}=2$ and $N_{\rm BDI}=1$ respectively. The phase boundaries are the two lines $ \Gamma_{z}-2 t_s= \pm \Delta$, on which the energy gap of $H_{\rm BDI}$ closes.

For a semi-infinite BDI class wire occupying the space with $y \geq 0$, the zero energy end states in the topological regime can be found in the continuum limit by solving $H_{\rm BDI}(k \rightarrow -i\partial_{y}) \gamma(y)=0$. In the regime with $N_{\rm BDI}=2$ where $2t_s-\Gamma_{z}> \Delta$, there are two solutions $\gamma_{1}(y)=[1,1,1,1](e^{-\lambda_{1+}y}-e^{-\lambda_{1-}y})$ and $\gamma_{2}(y)=i[1,1,-1,-1](e^{-\lambda_{2+}y}-e^{-\lambda_{2-}y})$. Here, $\lambda_{l,\pm} = \frac{t_{\rm so} \pm \sqrt{(t_{\rm so})^2+t_s (\Gamma -2t_s +(-1)^{l} \Delta)} }{t_s}$.  Note that the zero energy solutions satisfy the conditions $\gamma_{i} = \gamma_{i}^{\dagger}$, so that the end states are MFs. Moreover, under the time-reversal symmetry like operation $\mathcal{T}$, we have $\mathcal{T}\gamma_{i}\mathcal{T}^{-1}=\gamma_{i}$. As a result, the coupling between the two MF end states, which can be written as $i\gamma_{1}\gamma_{2}$, breaks the $\mathcal{T}$ symmetry. This term is not allowed so long as $\mathcal{T}$ is respected. Therefore, the two MF end states do not couple to each other, which is a feature of the BDI class topological superconductor. This is in sharp contrast to D-class topological superconductors where an even number of MFs can couple to each other and the MFs are lifted to finite energy.

It is interesting to note that as we approach the phase boundary between the $N_{\rm BDI}=2$ and $N_{\rm BDI}=1$ phases, where $\Gamma_{z} -2t_s = -\Delta $, we have $\lambda_{2 -} \to 0$ and $\gamma_{2}$ is no longer localized at the end of the wire. The process of approaching the phase boundary is depicted in Fig.1b and Fig.1c. In the regime where $N_{\rm BDI}=1$, only one MF end state $\gamma_{1}$ remains. In the regime where $N_{\rm BDI}=0$, there are no zero energy end state solutions.

For a long wire with length $L$ and neglecting the coupling between the left and right MFs, there are two more MF solutions $\gamma_{3}(y)=i[1,-1,-1,1](e^{\lambda_{3+}(y-L)}-e^{\lambda_{3-}(y-L)})$ and $\gamma_{4}(y)=[1,-1,1,-1](e^{\lambda_{4+}(y-L)}-e^{\lambda_{4-}(y-L)})$ as depicted in Fig.1b. It is worthwhile to note that the form of the MF wavefunctions are important for determining the transport properties of the superconductor as shown below. Moreover, we note that besides the set of symmetries discussed above, the Hamiltonian $H_{\rm BDI}(k)$ respects another time-reversal like symmetry $\mathcal{T}' =\mathcal{K}$ such that $\mathcal{T}' H_{\rm BDI}(k) {\mathcal{T}'}^{-1}=H_{\rm BDI}(-k)$. The four MF end states transform under $\mathcal{T'}$ as $\mathcal{T'}\gamma_{1/4}\mathcal{T'}^{-1}=\gamma_{1/4}$ and $\mathcal{T}'\gamma_{2/3}\mathcal{T'}^{-1}=-\gamma_{2/3}$.  
On the other hand, when the wire is finite, the MFs from the two ends of the wire can couple to each other. While the interaction terms $i\gamma_{1} \gamma_{4}$ and $i\gamma_{2} \gamma_{3}$ break the $\mathcal{T'}$ symmetry and are not allowed, the coupling terms $i\gamma_{1} \gamma_{3}$ and $i\gamma_{2} \gamma_{4}$ are allowed.

\subsection{Local Andreev Reflections}

It has been shown in previous works [\onlinecite{LLN, WADB}] that a single MF end state induces resonant local ARs at a normal lead/topological superconductor junction where an incoming electron is reflected as a hole in the same lead with probability of unity. The resonant local ARs result in zero bias conductance (ZBC) peaks of height $2e^2/h$ in transport measurements at zero temperature. It has also been shown that one dimensional DIII class topological superconductors, which respect time-reversal symmetry and particle-hole symmetry, support two MF end states at one end of the wire [\onlinecite{TK, Chris, Sho, Fan, XJ2, Liang}]. The two MF end states can induce a ZBC peak of height $4e^2/h$ [\onlinecite{Chris}]. Therefore, one may expect that the BDI class topological superconductor in the phases with $N_{\rm BDI}=1$ and $N_{\rm BDI}=2$ can both induce ZBC peaks in tunneling experiments. Surprisingly, we find that while the single MF end state in the $N_{\rm BDI}=1$ phase can induce ZBC peaks, the two MFs in the $N_{\rm BDI}=2$ phase completely suppress local ARs at zero bias and cause a conductance dip at low voltages.

The experimental setup for the BDI topological superconductor attached to a normal lead is depicted in Fig.2a. To calculate the tunneling spectroscopy of the BDI topological superconductor at different phases, we first write down a real space tight-binding model which corresponds to $H_{\rm BDI}(k)$ as described in the Methods section. A semi-infinite normal metal lead is attached to the left end of the topological superconductor. The zero temperature conductance of the normal metal/topological superconductor junction can be calculated from the reflection matrix $R_{\rm he}$ of the junction:
\begin{equation}
G = \frac{2e^2}{h} {\rm Tr} \left( R_{\rm he} R_{\rm he}^\dagger \right), \label{conductance}
\end{equation}
where $R_{\rm he}(E)_{ij}$ denotes the local AR amplitude of an electron with energy $E$ at channel $j$ to be reflected as a hole in channel $i$, which is calculated using the recursive Green's function approach [\onlinecite{Lee, Lee2, Sun,Jie}].

\begin{figure}
\begin{center}
\includegraphics[width=3.2in]{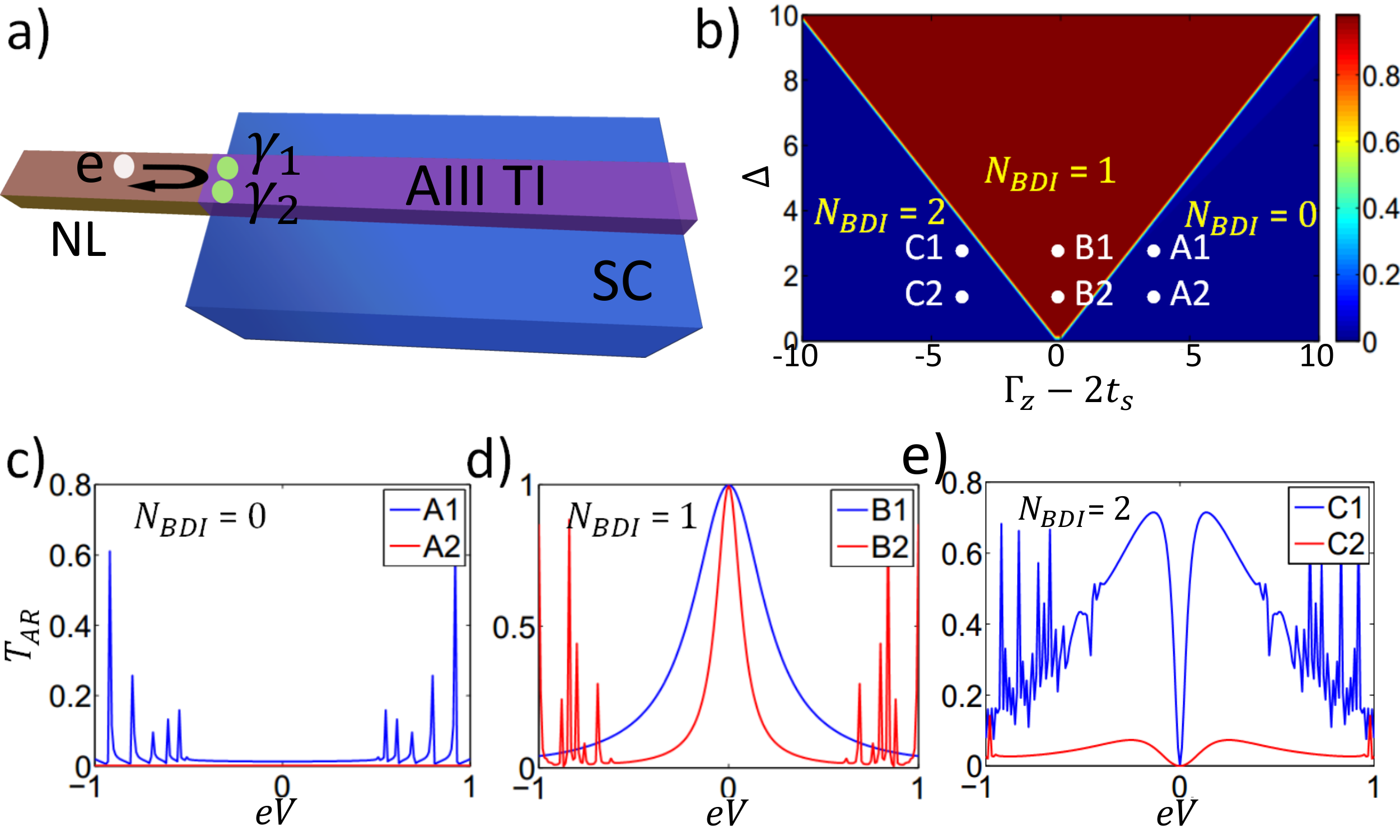}
\caption{ Transport properties of a normal lead/BDI class topological superconductor junction. (a) A normal lead is attached to one end of a semi-infinitely long BDI class topological superconductor in the $N_{\rm BDI}=2$ phase. At zero voltage bias, the electrons are totally reflected due to the destructive interference of the AR amplitudes induced by the two MF end states. (b) ZBC of the setup depicted in (a) as functions of $\Delta$ and $\Gamma_z$. It it evident that the ZBC is quantized at $2e^2/h$ in the $N_{\rm BDI}=1$ phase and zero otherwise. The details of the tight-binding model used is given in the Methods section. (c) to (e) The conductance as a function of voltages at points A1, A2, B1, B2, C1 and C2 denoted in (a) respectively. The conductance peaks at high voltages are due to bulk states and they appear only when the voltage bias is larger than the energy gap of the superconductor. }
\label{Fig2}
\end{center}
\end{figure}

The ZBC as a function of $\Delta$ and $\Gamma_z$ is shown in Fig.2b. As expected, in the phase with $N_{\rm BDI}=0$, the ZBC is strongly suppressed. When $N_{\rm BDI}=1$, the ZBC is quantized at $2\frac{e^2}{h}$ due to the MF induced resonant ARs [\onlinecite{LLN, WADB}]. Surprisingly, in the $N_{\rm BDI}=2$ phase, the ZBC is zero even though there are two zero energy MFs at the end of the topological superconductor. The conductance at finite voltages are shown in Fig.2c-e. It is evident from Fig.2e that there is a ZBC dip at the $N_{\rm BDI}=2$ phase instead of a ZBC peak. In the following, we construct an effective Hamiltonian of the normal lead/topological superconductor junction and show that the ZBC dip at $N_{\rm BDI}=2$ is due to destructive inference between the local AR amplitudes caused by the two MFs.

For voltage bias smaller than the pairing gap, we expect the transport properties of the junction to be described by an effective Hamiltonian
\begin{equation}
\begin{array}{lll}
H_{1\rm eff} &=& H_{\rm L} + H_{\rm LM} \\
H_{\rm L} &=& iv_{\rm F} \int_{-\infty}^{+\infty} \psi_{\rightarrow}^{\dagger}(y) \partial_{y} \psi_{\rightarrow}(y) dy \\
H_{\rm LM} &=& \omega_{1} \gamma_{1} [\psi_{\rightarrow}(0) - \psi_{\rightarrow}^{\dagger}(0)] + i \omega_{2} \gamma_{2} [\psi_{\rightarrow}(0) + \psi_{\rightarrow}^{\dagger}(0)].
\end{array}
\end{equation}
Here, $H_{\rm L}$ is the effective Hamiltonian for the left lead, $v_{\rm F}$ is the Fermi velocity of the lead. We note that, in general, one should consider a metal lead with electrons carrying spin pointing to the positive $x$-direction $\psi_{\rightarrow}$ and electrons carrying spin pointing to the negative $x$-direction $\psi_{\leftarrow}$. However, it can be shown that using the form of the wavefunctions of $\gamma_1$ and $\gamma_2$, that only $\psi_{\rightarrow}$  electrons can couple to the MF end states and $\psi_{\leftarrow}$ are decoupled from the superconductor. The form of the effective coupling term $H_{LM}$ is crucial for the study of the transport properties. The coupling between the left lead and the two MF end states of the topological superconductor is described by $H_{LM}$ and $\omega_i$ are the coupling amplitudes.

With $H_{1\rm eff}$, the scattering matrix can be easily calculated using the equation of motion approach [\onlinecite{LLN}] . It can be shown that the local AR amplitudes for an incoming electron with energy $E$ is $R_{\rm he}= -\omega_1^2/\zeta_1 + \omega_2^2/\zeta_2$, where $\zeta_{1,2} \equiv \omega_{1,2}^2+i E v_F/2 $. Therefore at $E=0$, $R_{\rm he}(E=0)=0$ as the two local AR amplitudes caused by the two MFs have opposite signs and they cancel each other out, as long as both $\omega_1$ and $\omega_2$ are finite. In other words, the suppression of the local ARs at zero bias is caused by the destructive interference of AR amplitudes caused by the two MF end states. This is in sharp contrast to the resonant ARs caused by a single MF end state in the D class case.

From the wavefunctions of the end states studied in Section IIA, we note that as $\Delta$ increases, $\gamma_1$ remains localized at the end and $\gamma_2$ merges into the bulk gradually. Then $\omega_2$ reduces to zero as $\Delta$ approaches the phase transition line $\Gamma_z - 2t_s = -\Delta$. Further increasing $\Delta$ would change the phase from $N_{\rm BDI}=2$ to the $N_{\rm BDI}=1$ phase. When $\omega_2=0 $ in the $N_{\rm BDI}=1$ phase, we have $|R_{\rm he}(E=0)|=1$ and the resulting ZBC is $2e^2/h$ according to Eq.\ref{conductance} as expected [\onlinecite{LLN, WADB}].

To understand the transport properties at finite voltages, we note that when $\omega_2 \ll \omega_1$, the local AR amplitudes become significant when the energy of the incoming electrons reaches $|E| \approx 2|\omega_2^2|/v_{F}$. As a result, the width of the ZBC dip becomes narrower as $\Delta$ increases, as shown in Fig.2e, and the ZBC dip disappears when $\omega_2$ goes to zero.

\subsection{Resonant crossed Andreev reflections}

In the above sections, it is shown that local AR processes are suppressed at a normal lead/topological superconductor junction for the $N_{\rm BDI}=2$ phase. Due to the suppression of the local AR amplitudes and the conservation of probability, we expect that other tunneling processes can become more important. In this section, we show that the two MF end states in the $N_{\rm BDI}=2$ phase can strongly enhance the crossed AR processes in a normal lead/topological superconductor/normal lead junction, provided that the length of the superconducting wire is comparable to the localization lengths of the MF end states such that the MFs from the two ends can couple to each other. In a crossed AR process, an electron from one lead is reflected as a hole in the other lead. As a result, two electrons from the two leads form a Cooper pair and get injected into the superconductor, as depicted in Fig.3a.

To calculate the transport properties of the superconductor, we attach two normal leads to the superconductor as depicted in Fig.3a. The superconductor is
described by a tight-binding model presented in the Methods section. The length of the superconductor is $L=20a$ which is comparable with the localization length of the MF end states. Here, $a$ is the lattice constant of the tight-binding model and the parameters of the model is given in the Methods section. Focusing on the transport properties of the left normal lead, the local AR amplitudes, the crossed AR amplitudes, the elastic electron co-tunneling amplitudes and the electron normal reflection amplitudes for the three different phases at zero bias are shown in Fig.3. It is surprising that, in the $N_{\rm BDI}=2$ phase, there are parameter regimes where the crossed AR amplitude is unity. When this happens, all other tunnelling amplitudes for the $\psi_{\leftarrow}$ electrons, including the elastic co-tunneling amplitudes for which electrons tunnel directly from the left lead to the right
lead, vanish.

\begin{figure}
\begin{center}
\includegraphics[width=3.2in]{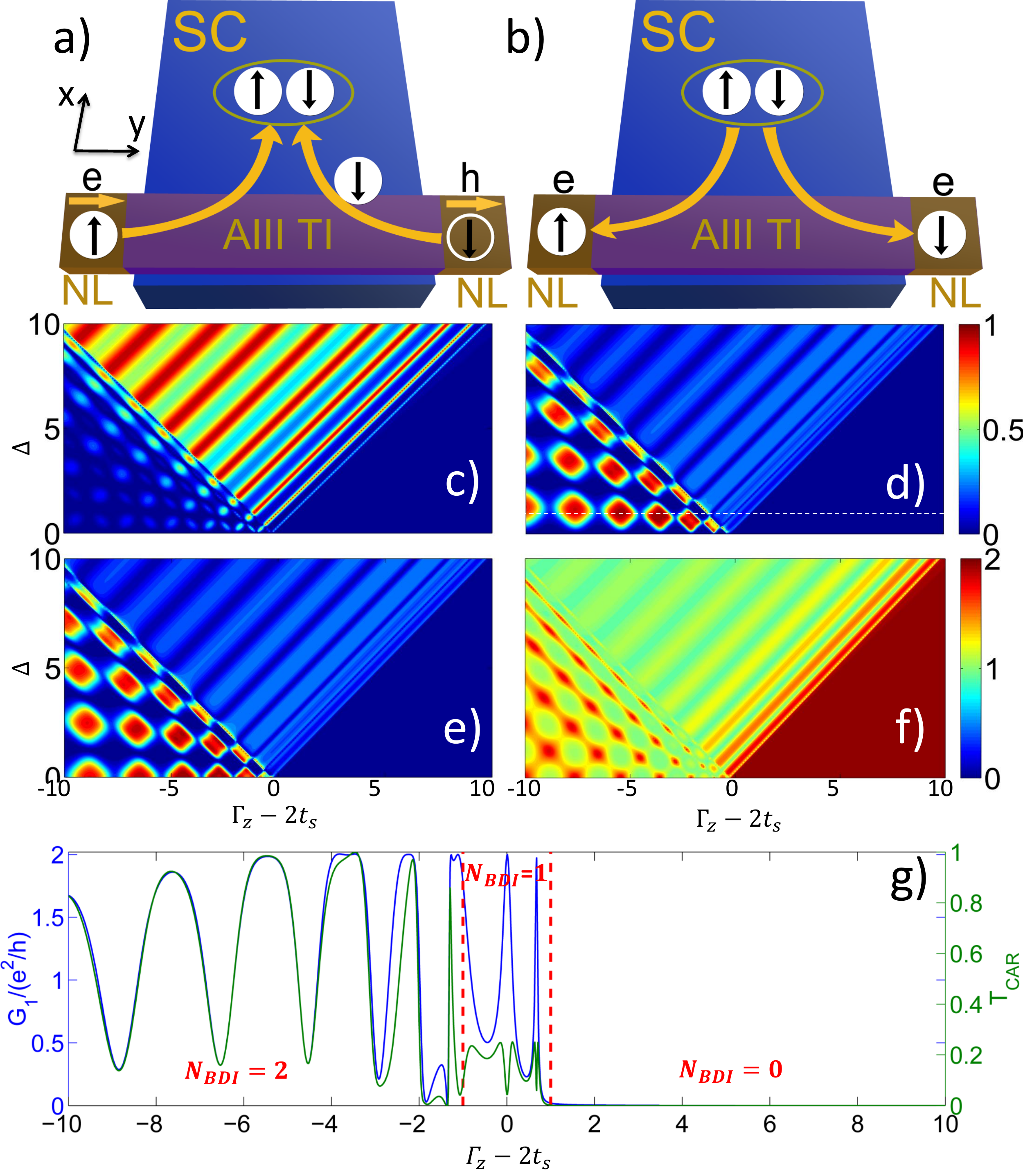}
\caption{ Transport properties of a normal lead/topological superconductor/normal lead junction. (a) Two normal leads are attached to the two ends of a wire with finite length. In a crossed AR process, an electron from the left lead is reflected as a hole in the right lead. As a result, two electrons are injected into the superconductor to form a Cooper pair. For our setup, all the electrons on the left (right) lead have spin pointing to the positive (negative) $x$-direction. (b) In a Cooper pair splitting process, a Cooper pair is split into two electrons with opposite spins. One electron is injected into each lead. (c) Local AR for a short wire with length $L=20a$. In this case, the local ARs are still strongly suppressed but they can deviate from zero. d) Crossed AR amplitudes at zero bias of the wire. The crossed AR amplitudes can be close to one in the $N_{\rm BDI}=2$ phase for a large phase space. e) Elastic co-tunneling amplitudes at zero bias. f) Normal reflection amplitudes at zero bias. It is important to note that the normal reflection amplitudes have a minimal value of one. This is due to the fact that one of the spin channels of the normal lead is completely decoupled from the superconductor and all the electrons of that spin channel are reflected. g) The ZBC ( blue line) and the crossed AR amplitude (green line) as functions of $\Gamma_z$ for parameters denoted by the horizontal dashed line in d). The vertical red dashed lines separate the three phases with different $N_{\rm BDI}$.}
\label{Fig3}
\end{center}
\end{figure}

On the other hand, crossed AR amplitudes in the $N_{\rm BDI}=1$ phase have similar properties as the cases of D class topological superconductors [\onlinecite{Nilsson, Jie}]. In this phase, there are regimes where local AR processes are suppressed and the crossed AR processes dominate. However, crossed AR amplitudes are always equal to the elastic co-tunneling processes in the $N_{\rm BDI}=1$ phase [\onlinecite{Nilsson,Jie}]. Therefore, the crossed AR cannot reach unity.  As shown in Fig.3e, the maximal crossed AR amplitude is in general much smaller than unity in the $N_{\rm BDI}=1$ phase. Therefore, the possibility of inducing resonant crossed ARs is a unique signature of the $N_{\rm BDI}=2$ phase.

To understand the numerical results, we expect the transport properties for voltage bias smaller than the superconducting pairing gap to be well described by an effective Hamiltonian which includes the coupling between the MFs with the two leads as well as the coupling among the four MF end states. The Hamiltonian reads:
\begin{equation}
\begin{array}{lll}
H_{2\rm eff} &= & H_{\rm L} + H_{\rm R}+ H_{\rm M}+H_{\rm LM} +H_{\rm RM} \\
H_{R} &= & iv_{F} \int_{-\infty}^{+\infty} \psi_{\leftarrow}^{\dagger}(y) \partial_{y} \psi_{\leftarrow}(y) dy \\
H_{M} &= & i E_{13}\gamma_{1} \gamma_{3} + i E_{24} \gamma_{2} \gamma_{4} \\
H_{RM} &=& i \omega_{3} \gamma_{3} [\psi_{\leftarrow}(0) + \psi_{\leftarrow}^{\dagger}(0)] + \omega_{4} \gamma_{4} [\psi_{\leftarrow}(0) - \psi_{\leftarrow}^{\dagger}(0)].
\end{array}
\end{equation}
The Hamiltonian of the left lead $H_{\rm L}$ and the coupling between the left lead and the MFs $H_{\rm LM}$ have been discussed above.  Here, $H_{\rm R}$ describes the right normal lead and $\psi_{\leftarrow}$ denotes an annihilation operator of an electron with spin pointing to the negative $x$-direction. It is important to note that for the right lead, only electrons with spin polarized along the negative $x$-direction are coupled to the MFs due to the form of the MF wavefunctions $\gamma_3$ and $\gamma_4$. $H_{\rm RM}$ describes the coupling between the right lead and the MFs. The coupling between the four MF end states is described by $H_{\rm M}$, where $E_{13}$ and $E_{24}$ are real numbers denoting the coupling strength between the MFs from the opposite ends of the wire. As discussed above, the coupling terms such as $i \gamma_1 \gamma_4$ and $i\gamma_2 \gamma_3$ are not allowed by symmetry.

For the effective Hamiltonian $H_{2\rm eff}$, the scattering matrix can be found and the crossed AR amplitudes from one lead to another lead at $E=0$ is $-\omega_{1} \omega_{3} E_{13}v_{F}/(E_{13}^2v_{F}^2+\omega_{1}^2 \omega_{3}^2) - \omega_{2} \omega_{4} E_{24}v_{\rm F}/(E_{24}^2v_{\rm F}^2+\omega_{2}^2 \omega_{4}^2)$. Crossed AR processes are depicted in Fig.3a. When both the conditions $E_{13}/v_{\rm F}=\omega_1 \omega_3$ and $E_{24}/v_{\rm F}=\omega_2 \omega_4$ are satisfied, the crossed AR amplitude is unity and all other tunneling amplitudes are zero. We call this phenomenon resonant crossed ARs. As shown in Fig.3d, there is a sizeable phase space in which the crossed AR amplitudes are close to one. The oscillating behavior of the tunneling amplitudes in the phases with MFs is due to the fact that the coupling strengths of the MFs oscillate as a function of $\Delta$ and $\Gamma_z$ [\onlinecite{Jie}].

As depicted in Fig.3b, the reverse processes of the crossed ARs are the Cooper pair splitting processes. When a current is driven from the superconductor to the two leads, a Cooper pair from the superconductor can be split into two spatially separated but
correlated electrons and one electron is injected into each of the two leads. In the language of scattering matrix, the Cooper pair splitting amplitude is equivalent to the amplitude for an incoming hole from the left lead to be reflected as an electron in the right lead. One can show that the Cooper pair splitting amplitude equals the crossed AR amplitude. As a result, when a current is driven from the superconductor to the leads, we can have resonant Cooper pair splitting.

Remarkably, for the left lead, only electrons with spin pointing to the positive $x$-direction are coupled to the superconductor and for the right lead, only electrons with spin pointing to the  negative $x$-direction are coupled to the superconductor due to symmetry constraints. Therefore, the current of the left (right) lead is spin-polarized to the positive (negative) $x$-direction. Moreover, due to the resonant crossed Andreev reflections, the conductance of each normal lead is $G=2e^2/h$ and the current is spin-polarized. The ZBC of the left lead, with parameters corresponding to the horizontal dashed line in Fig.3d, is shown in Fig.3g.

In Fig.3g, the ZBC is denoted by the blue line and the crossed AR amplitudes are denoted by the green line. As $\Delta$ is fixed and $\Gamma_z$ increases, all the three phases with $N_{\rm BDI}=2$, $ 1$ and $0$ can be reached. In the $N_{\rm BDI}=2$ phase, it is clear that the conductance is almost solely determined by the crossed AR amplitude as the local AR amplitudes are strongly suppressed as shown in Fig.3c. When the crossed AR amplitude approaches unity, the ZBC approaches $2e^2/h$. In the $N_{\rm BDI}=1$ phase, the conductance can reach $2e^2/h$ due to local ARs. In the $N_{\rm BDI}=0$ phase, the ZBC goes to zero. Since the currents out of the left and right leads are spin-polarized, and the fact that there are no spin-orbit coupling in the normal lead, the normal lead can sustain a spin current. Therefore, the BDI class topological superconductor in the $N_{\rm BDI}=2$ phase can be a novel source of conserved spin currents for spintronic applications.

\subsection{Realistic Cooper pair splitters}
In this section, we point out that the anomalous transport properties of BDI class topological superconductor discussed above can be experimentally realized using anomalous Hall insulators in proximity to an $s$-wave superconductor.

A quantum anomalous Hall insulator (QAHI) is an insulator with gapless chiral fermionic edge states in the absence of an external magnetic field, which has been experimentally discovered recently [\onlinecite{Chang}]. Interestingly, it was shown by Qi et al. [\onlinecite{Qi}] that in proximity with an $s$-wave superconductor, a QAHI can be turned into a topological superconductor which supports one or two branches of chiral MF edge states, as depicted in Fig.4a. The topological superconducting phases can be classified by Chern numbers $N_{\rm Chern}$ with $N_{\rm Chern}$ denoting the number of branches of MF edge states. The Hamiltonian of a QAHI in the presence of superconducting pairing and in the Nambu basis $\{\phi_{\bf k\uparrow}, \phi_{\bf k\downarrow}, \phi^\dagger_{\bf -k\uparrow}, \phi^\dagger_{\bf -k\downarrow}\}$ can be written as:
\begin{equation}
\begin{array}{ll}
H_{\rm QAHI+\rm S}({\bf k}) =& [{\Gamma_z}'-2t'_s (\cos k_x+\cos k_y)]\tau_z \sigma_z \\
&+ 2 t'_{\rm so} \left( \sin k_x \tau_0 \sigma_x + \sin k_y  \tau_z \sigma_y \right)+\Delta \tau_y \sigma_y.
\end{array}
\end{equation}
Here, $\Gamma_{z}'$, $t'_{\rm s}$ and $t'_{\rm so}$ are real numbers characterizing the model [\onlinecite{Qi}]. For general momentum ${\bf k}$, the Hamiltonian is in the D class which respects only the particle-hole symmetry. The time-reversal like symmetries $\mathcal{T}$ and $\mathcal{T'}$ are broken by the $\sin k_x \tau_0 \sigma_x$ term. However, for $k_x=0$, $H_{\rm QAHI+ \rm S}$ is equivalent to $H_{\rm BDI}$ in Eq.2. As a result, the $k_x=0$ component of $H_{\rm QAHI+\rm S}$ is a BDI class topological superconductor. Moreover, the $N_{\rm Chern}=1$ ($N_{\rm Chern}=2$) phase in the quantum Anomalous Hall system corresponds to the $N_{\rm BDI}=1$ ($N_{\rm BDI}=2$) phase of the BDI class topological superconductor.

\begin{figure}
\begin{center}
\includegraphics[width=3.2in]{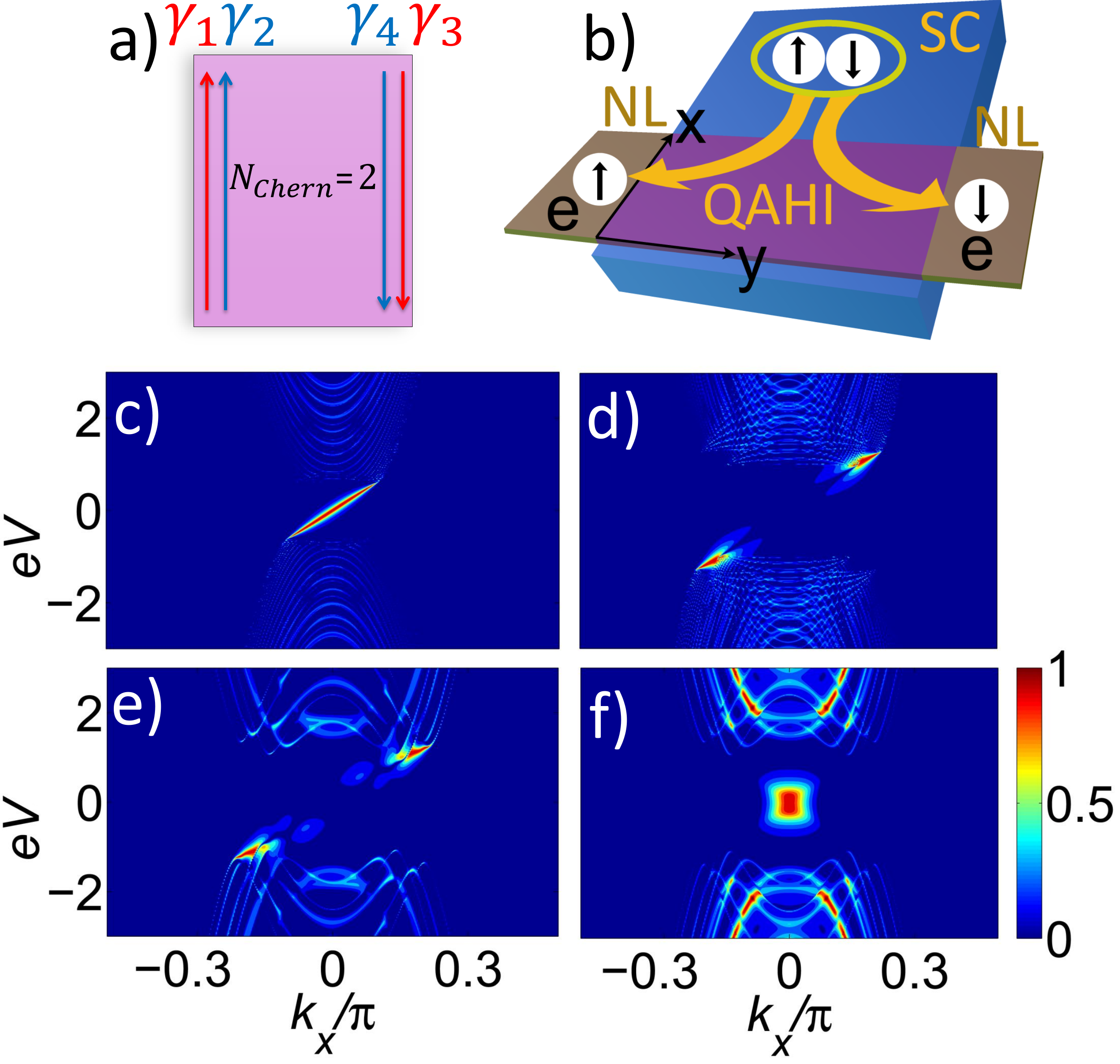}
\caption{Transport properties of a QAHI in proximity to an $s$-wave superconductor. (a) In the $N_{\rm Chern}=2$ phase of a QAHI in proximity to an $s$-wave superconductor, there are two branches of MF edge states localized at the edge of the system. The MF edge states are denoted by $\gamma_i$. (b) Two normal leads are attached to the edges of a QAHI. A Cooper pair is split into two electrons. The electrons injected to the left lead and the right lead have definite spin polarizations. This results in spin-polarized currents leaving the normal leads. (c) The momentum resolved conductance from the normal lead to a wide strip of superconducting QAHI in the $N_{\rm Chern}=1$ phase. The tight-binding model used is described in the Methods section. The width of the QAHI is $L_{y}=200a$. Periodic boundary conditions in $x$ direction is assumed. The strong local AR amplitudes at energy within the gap of the superconducting QAHI is due to the chiral MF edge state. (d) The local AR amplitudes in the $N_{\rm Chern}=2$ phase with $L_{y}=200a$. The local AR amplitudes are strongly suppressed at low voltages even in the presence of two chiral MF edge states as depicted in (a). (e) The momentum resolved local AR amplitudes in the $N_{\rm Chern}=2$ phase for a narrow strip of superconducting QAHI with $L_{y}=20a$. The distance between the two edges in the $y$-direction is comparable to the localization length of the chiral edge states. (f) The momentum resolved crossed AR amplitudes of a narrow strip of superconducting QAHI in the $N_{\rm Chern}=2$ phase with $L_y=20a$. The crossed AR amplitudes can be close to one at low voltages near $k_x=0$. }
\label{Fig4}
\end{center}
\end{figure}

A strip of QAHI in proximity to a superconductor and attached to two metal leads is depicted in Fig.4b. The tight-binding model used to describe
$H_{\rm QAHI+\rm S}$ is presented in the Methods section. The momentum resolved local AR amplitudes from the left normal lead to the QAHI in the
$N_{\rm Chern}=1$ and $N_{\rm Chern}=2$ phases are shown in Fig.4c and Fig.4d respectively. The width of the QAHI in this case is $L_{y}=200a$,
which is much longer than the localization length of the MF edge states. Focusing on the transport properties at $k_x=0$, we note that the local AR resonates at zero bias for the $N_{\rm Chern}=1$ phase but is suppressed for the $N_{Chern}=2$ phase.
Similar results were obtained by Ii et al. [\onlinecite{Ii}] while the reasons of
the transport anomalies were not given. By establishing the correspondence between BDI class topological superconductor and $H_{\rm QAHI+\rm S}$, we have shown with the effective tunneling Hamiltonian approach that the suppression of the local AR at the $N_{\rm Chern}=2$ phase is a consequence of the destructive interference of the AR amplitudes induced by the two MFs with $k_x=0$ at the edge of the QAHI.

Due to the strong suppression of the local AR amplitudes near $k_x=0$ in the $N_{\rm Chern}=2$ phase, we expect that the crossed AR amplitudes can be enhanced near $k_{x}=0$ when the width of the QAHI is reduced. The local AR and the crossed AR amplitudes for a narrow strip of QAHI with width $L_y=20a$ is presented in Fig.4e and Fig.4f respectively. From Fig.4f, it is
shown that the crossed AR amplitudes can reach almost unity for $k_x \approx 0$ at low voltage bias. At the same time, the local AR amplitudes in the $N_{\rm Chern}=2$ phase is strongly suppressed for this narrow strip of QAHI. As a result, similar to the case of the BDI topological superconductor in the $N_{\rm BDI}=2$ phase, when a current is driven from the superconductor to the lead, the QAHI can split the Cooper pairs effectively and result in correlated spin-polarized currents leaving the two normal leads.

\section{Discussion}
In this work, we show that the BDI class topological superconductor in the $N_{\rm BDI}=2$ phase can be used as an efficient Cooper pair splitter whereby the Cooper pairs can be split into two streams of spin-polarized currents.  Two important results are used to reach these conclusions, namely, the suppression of local ARs and the fact that the MF end states only couple to electrons with fixed spin polarizations of the leads. In this section, we argue that these results can be understood easily in the regime with small paring amplitudes.

First, since the local ARs compete with crossed ARs due to conservation of probability, the MFs should not induce strong local ARs, as in the case of D class topological superconductors. Otherwise, the crossed AR amplitudes would be small. For a BDI class topological superconductor in the $N_{\rm BDI}=2$ phase obtained by inducing superconductivity on a AIII class topological insulator with fermionic end states, the suppression of local ARs at zero bias is indeed quite natural.

Suppose that the AIII class topological insulator is in the non-trivial phase with a fermionic end state, adding a small superconducting pairing term does not close the energy gap and there is no topological phase transition. In this case, the fermionic end state can be regarded as two MF end states. Therefore, we have a BDI class topological superconductor with $N_{\rm BDI}=2$. However, when the pairing terms are zero, there cannot be any local ARs since the system is simply an insulator. Consequently, one may expect that the local AR amplitudes are strongly suppressed when the pairing amplitudes are small. The suppression of the local AR amplitudes opens up the possibility for the crossed AR amplitudes to be enhanced in the presence of finite $\Delta$.

It is important to note that the suppression of local ARs in the $N_{\rm BDI}=2$ phase does not contradict the results of Diez et al. [\onlinecite{Diez}] who predicted that the conductance at zero bias should be $N_{\rm BDI} \frac{2e^2}{h}$ at a normal lead/BDI topological superconductor junction. The reason is that the results obtained in Ref.[\onlinecite{Diez}] would apply only if $N_{\rm BDI}$ is calculated using the chiral symmetry $\mathcal{C'}=\mathcal{T'} \mathcal{P}$ where $\mathcal{P}$ is the particle-hole symmetry operator and $\mathcal{T'}=\mathcal{K}$ and $\mathcal{K}$ is the complex conjugate operator. This chiral symmetry $\mathcal{C'}$ is respected by $H_{\rm BDI}$ of Eq.2. Using this set of symmetries, one would find that the topological invariant $N_{\rm BDI}'$ equals to zero in the parameter regimes where $N_{\rm BDI}=2$ and $N_{\rm BDI}=0$. Moreover, $N_{\rm BDI}'=1$ in the regime where $N_{\rm BDI}=1$. Therefore, the ZBC should be zero in both the $N_{\rm BDI}=0$ and $N_{\rm BDI}=2$ phases. This is consistent with the results by Diez et al. However, the symmetry arguments alone are not enough to understand the conductance at finite voltages.  In short, the $N_{\rm BDI}=2$ topological phase in this work is different from the $N_{\rm BDI}'=2$ phases found in previous works [\onlinecite{Tewari, Chris}] as a different set of symmetry operators were used to calculate the topological invariants. The suppression of local ARs caused by usual fermionic Andreev bound states was also studied by Ioselevich and Feigel’man [\onlinecite{IF}]. However, resonant crossed ARs cannot happen in trivial superconductors due to the lack of symmetry constraints to restrict the form of the interactions among different Andreev bound states.

Second, by definition, a AIII class topological insulator respects a chiral symmetry. As a result, a non-degenerate zero energy fermionic end state at one end of the system has to be an eigenstate of the chiral symmetry operator.  For the AIII class model used in Eq.1, the chiral operator is $\sigma_{x}$. Therefore, the two end states at opposite ends of the wire are eigenstates of $\sigma_x$ with opposite eigenvalues [\onlinecite{XJ}]. If there are no spin flip terms in the leads, the end states can only couple to electrons which have the same spin as the end states. Using the form of the MF wavefucntions, one can show that this is true even in the presence of the pairing terms. As a result, in the effective Hamiltonian $H_{2eff}$, one can regard the left and the right normal leads as having opposite spin. This result is important for obtaining spin-polarized currents in the leads by splitting Cooper pairs.

Moreover, experiments on the efficient splitting of Cooper pairs using Coulomb blockade effect [\onlinecite{Loss}] have been reported [\onlinecite{SBS,Das}]. However, the currents leaving the superconductors are not spin-polarized and it is not known whether the electrons on different leads are correlated [\onlinecite{Bernd}]. Therefore, being able to generate correlated spin currents by splitting Cooper pairs is a very unique property of the BDI class topological superconductor.

Finally, we discuss the stability of the topological phases which support two zero energy Majorana modes on each edge of the system. As discussed above, the $N_{\rm BDI}=2$ phase of the BDI class topological superconductor is protected by the chiral symmetry $\mathcal{C}=\sigma_{x}\tau_{0}$. Therefore, terms such as $\sigma_x \tau_{z}$ can break the chiral symmetry and make the superconductor topologically trivial. However, the $N_{Chern}=2$ phase of the superconducting QAHI is in the D-class [\onlinecite{SRFL}] and the zero energy Majorana modes on the edge are robust against perturbations as long as the bulk gap is not closed. Therefore, one can always extract an effective one dimensional Hamiltonian from the two dimensional superconducting QAHI which supports two Majorana modes at each edge. The coupling of this one dimensional Hamiltonian with normal leads can be described by Eq.4. Therefore, the suppression of local Andreev reflections and the appearance of almost resonant crossed Andreev reflections in the QAHI case are robust against perturbations.

\section{Methods}
\subsection{Tight-binding models}
For the calculations of the momentum resolved transport properties of the QAHI with superconducting pairing terms, we apply periodic boundary conditions in the $x$-direction and open boundary conditions in the $y$-direction. Spinful normal leads are attached to the two edges parallel to the $x$-directions. The tight-binding model for a strip of QAHI with superconducting pairing terms can be written as:
\begin{widetext}
\begin{equation}
\begin{array}{lll}
H_{QAHI+S}(k_x)&=&\sum_{i} [ -t'_s(c^\dagger_{i,k_x,\uparrow}c_{i+1,k_x,\uparrow} -c^\dagger_{i,k_x,\downarrow}c_{i+1,k_x,\downarrow})
+t'_{\rm so}(c^\dagger_{i,k_x,\uparrow}c_{i+1,k_x,\downarrow}-c^\dagger_{i,k_x,\uparrow}c_{i-1,k_x,\downarrow})]+H.c\notag\\
&&+\sum_i [({\Gamma_z}'-2t'_s\cos k_x )(c^\dagger_{i,k_x,\uparrow}c_{i,k_x,\uparrow}-c^\dagger_{i,k_x,\downarrow}c_{i,k_x,\downarrow}
+2t'_{\rm so}\sin k_x ( c^\dagger_{i,k_x,\uparrow}c_{i,k_x,\downarrow}+c^\dagger_{i,k_x,\downarrow}c_{i,k_x,\uparrow} )\notag\\
& &+ \Delta (c^\dagger_{i,k_x,\uparrow}c^\dagger_{i,-k_x,\downarrow}+h.c.)].
\end{array}
\end{equation}
\end{widetext}

Here, $c_{i, k_x, \uparrow}$ ($c_{i,k_x, \downarrow} $) denotes an electron operator at site $i$ along the $y$-direction and has momentum quantum number $k_{x}$ along the $x$-direction and spin up ( spin down ) with respect to the $z$-direction. In all the figures in Fig.4, the parameters are: $t'_{\rm so}=1$, $t'_{\rm s}=10$. For Fig.4c and Fig.4e, ${\Gamma_{z}}'=40$ and $\Delta=1$ so that the system is in the $N_{Chern}=1$ phase. In Fig.4d and Fig.4f, ${\Gamma_{z}}'=36.4$ and $\Delta=1$ so that the system is in the $N_{ \rm Chern}=2$ phase.

The same tight-binding model $H_{QAHI+S}(k_{x})$, with $k_x=0$, can be used to describe the BDI class topological superconductor $H_{\rm BDI}(k)$ of Eq.2 with parameters $t'_{\rm so} = t_{\rm so}$, $t'_{\rm s}=t_{\rm s}$ and ${\Gamma_{z}}'=\Gamma_{z} + 2 t_{\rm s}$. In Fig.2 and Fig.3, $t_{\rm s}=10$ and the number of sites in the $y$-direction is $L=200a$ and $L=20a$ respectively where $a$ is the lattice constant.

{\bf Acknowledgement}\\
The authors thank Masatoshi Sato, Keiji Yada and Yi Zhou for discussion. KTL is indebted to Tai Kai Ng for insightful discussions and his encouragements throughout this project. KTL acknowledges the support of HKRGC through Grant 605512, Grant 602813 and HKUST3/CRF09.  YT thanks the support of MEXT of Japan through Grant 22103005 and Grant 2065403. After the submission of this work, we note that there is an independent work by A. Yamakage and M. Sato on the study of the suppression of the local ARs of the $N_{\rm Chern}=2$ phase [\onlinecite{Yamakage}]. 

{\bf Author contributions} \\
J.J.H. and J.W. are involved in the analytic and numerical calculations. T.P.C. and X.J.L. are involved in analyzing the models. Y.T. and K.T.L initiated and supervised the project. K.T. L. conceived the ideas of this paper and prepared the manuscript with contributions from all the authors.\\

{\bf Competing financial interests}: The authors declare no competing financial interests.

\end{document}